\documentstyle[amssymb,multicol,epsf,aps]{revtex}
\def\dbarrm {{\mathchar'26\mkern-11mu{\rm d}}}                        %

\newcommand{\BEQ}{\begin{equation}}
\newcommand{\EEQ}{\end{equation}}
\newcommand{\BEA}{\begin{eqnarray}}
\newcommand{\EEA}{\end{eqnarray}}
\renewcommand{\H}{{\cal H}}

\renewcommand{\d}{{\rm d }}

\newcommand{\half}{\frac{1}{2}}

\newcommand{\hT}{{T/2}}

\renewcommand{\S}{S_{\rm ep}}
\newcommand{\p}{\partial}
\newcommand{\minfty}{{-\infty}}

\newcommand{\I}{{\cal I}}

\newcommand{\Kt}{{\tilde K}}

\newcommand{\Ht}{{\tilde H}}

\begin{document}
\draft
\title{Solvable model for the standard folklore of the glassy state}
\author{Theo~M.~Nieuwenhuizen}
\address{Department of Physics and Astronomy, University of Amsterdam
\\ Valckenierstraat 65, 1018 XE Amsterdam, The Netherlands}
\date{Revised, September 28, 1999}% ; printout: \today}
\maketitle
\begin{abstract}
A  model system with fast and slow processes is introduced.
After integrating out the fast ones, 
the considered dynamics of the slow variables is exactly solvable.
In statics the system undergoes a Kauzmann transition to a glassy state.
The relaxation time obeys a generalized 
Vogel-Fulcher-Tammann-Hesse law.
The aging dynamics on the approach to and below the Kauzmann temperature 
is analyzed; it has logarithmic behavior.
The structure of the results could be general, as they satisfy
laws of thermodynamics far from equilibrium. 
The original VFTH law is on the border-line between the regime 
where only the effective temperature of the slow modes is needed,
and the regime where also an effective field occurs.
The production rates of entropy and heat are calculated.
\end{abstract}
\begin{multicols}{2}
 
Half a century ago Kauzmann pointed at the 
``entropy crisis'' in glass forming liquids:
The naive extrapolation of the entropy of an undercooled liquid 
goes below the entropy of the crystalline state, which is 
physically incorrect~\cite{Kauzmann}.  
It is often assumed  that a thermodynamic phase transition takes place 
around the  ``Kauzmann temperature'' $T_K$, characterized by a
vanishing configurational entropy.
This prediction is very difficult to test experimentally, 
since the relaxation time may be very long.
Experimental data for glass forming liquids
are often fit to a Vogel-Fulcher-Tammann-Hesse (VFTH)~\cite{VTF} 
behavior $\tau_{\rm eq}$ $\sim$
$\exp [A^\gamma/(T-T_0)^{\gamma}]$. An entropic argument for the
 standard choice, $\gamma=1$, was given by Adam and
Gibbs~\cite{AdamGibbs}. A field-theoretic approach by 
Parisi~\cite{ParisiVF} leads to $\gamma> 1$. 
These approaches share the commonly accepted point that $T_0$ and $T_K$
coincide. In practice exponents $\gamma\neq 1$ are compatible 
with data; this merely affects the width of the fitting interval.
Here we shall investigate universal aspects of this standard picture.

Very little is known about dynamics below the Kauzmann temperature.
One simple, intriguing idea comes to the mind. Consider a glassy
system that has aged a long time $t$. Can one say that all processes
with equilibration time much less than $t$ are in equilibrium,
while the ones with timescale much larger than $t$ are still
quenched, leaving the processes with timescale of order $t$ as the only
interesting ones? Such an approach would be too naive for continuous 
phase transitions, where the timescale diverges
algebraically at the critical point. 
However, in glasses there is an exponential divergence,
and this might induce an asymptotic decoupling of the time-decades.
Indeed, it  occurs in the glassy regime of  
models with an Arrhenius (rather than VFTH) law,  
like the ferromagnetic Ising chain ~\cite{Nthermo}, 
harmonic oscillator and spherical spin models~\cite{Nhammer}.
This asymptotic decoupling could be the basic ground for 
a generalization of equilibrium thermodynamics to systems 
far from equilibrium~\cite{Nhammer}.
That approach involved systems where one extra variable was
needed to describe the non-equilibrium physics, namely the
effective temperature. In the present work we shall
encounter situations, where also
an effective field is needed to describe the
two observable quantities, energy and magnetization out of equilibrium.
That  case is of principle interest for glass forming
liquids, as it was found experimentally that at given $T,p$ the same 
volume can be reached via different histories.
As they then lead to different futures,
the glassy state of these systems
cannot be coded using one extra variable~\cite{McKenna}.

In order to gain insight in properties of the standard picture,
we introduce a simple model, with solvable dynamics. In the
long time aging regime it leads to first order differential  
equations. We thus can study aging
 on the approach to and below the glass transition.
Even though the physics of the model is quite simple, we shall 
set out to discover general aspects of the results by
 phrasing them in a thermodynamic language.

The model contains a set of $N$ spins $S_i$, coupled to a slowly 
changing background field, that we model by a set of $N$ harmonic 
oscillators $x_i$. All variables are only coupled locally,  
as is described by  the Hamiltonian
\BEQ \label{H1=}
\H=\half K\sum_ix_i^2-H\sum_ix_i-J\sum_i x_iS_i-L\sum_iS_i
\EEQ
The spins have no fixed length, but satisfy the 
spherical constraint $\sum_i S_i^2=N$. 
In order to model the $\beta$-processes of a glass,
we assume that the spins $S_i$ move fast, and we
shall consider only the time regime where they
are at thermodynamic equilibrium.
To model $\alpha$-processes, we shall
consider a dynamics where the $x_i$ may fall out of equilibrium.
Integrating out the $S_i$ replaces the spin-part of the energy (\ref{H1=}) 
by its free energy, which will act as an {\it effective} 
Hamiltonian for the slowly moving variables $x_i$,
\BEQ 
\H_{\rm eff}=U-TS_{\rm ep}
\EEQ
Here $U=Nu$ is the internal energy given by
 \BEQ \label{u=}
u=\frac{U}{N}=\half Km_2-Hm_1-w+\frac{T}{2}, \EEQ
It depends on the two lowest moments of the $x_i$,
$m_k=(1/N)M_k=(1/N)\sum_i x_i^k$,
and $w$ is the short hand
\BEQ
w=\sqrt{J^2m_2+2JLm_1+ L^2+{\small \frac{1}{4}}T^2},
\EEQ
$\S$ is the entropy of the {\bf e}quilibrium {\bf p}rocesses, 
i.e., of the spins,
\BEQ \label{Sep=}
\S= \frac{1}{2}N-\frac{1}{2}N\ln[\beta (w+\half T)]
\EEQ
Being interested in the long-time aging behavior,
we may restrict ourselves to the time-regime
where the dynamics of the $x_i$ can be
described by a Monte Carlo dynamics. 
We must assume parallel updating, 
as sequential updating would not produce a long timescale in
our system without two-particle interactions. We are thus considering 
a simple example of models typically studied numerically. 
Since it can even be solved analytically, we shall gain qualitative 
insight in the long time regime. 

The rules for the MC dynamics are standard, and lead to
evolution equations for $m_{1,2}$ that have the form
\BEA \label{dotm1=} 
\dot m_1&=& \int_\minfty^\infty \d x\,y_1(x)p(x|m_{1,2}) 
W(\beta x) \\ \dot m_2&=& 2 \int_\minfty^\infty  
\d x\, \frac{x+\Ht y_1(x)}{\Kt} p(x|m_{1,2}) W(\beta x)
\label{dotm2=}\EEA
Let us explain how this comes about.
For one MC step one shifts 
all $x_i$ by an amount $r_i/\sqrt{N}$ 
with the $r_i$ drawn from a symmetric
Gaussian with variance $\Delta^2$. 
This updates $ M_{1,2}'=M_{1,2}+\delta M_{1,2}$.
According to the Metropolis algorithm,
the move is accepted with probability 
$W(\beta x)=\exp(-\beta x)$ if
the change in effective energy, $x=\H_{\rm eff}'-\H_{\rm eff}$,
 is positive; if $x$ is negative 
the move is always accepted ($W(\beta x)=1$). Such dynamical 
problems are well understood~\cite{BPR}~\cite{Nlongthermo}. 
The essential ingredient is the thermally averaged transition 
probability for a move characterized by $x$ and $\delta M_1$. 
It is the product of two factors: 
the first is a  Gaussian in $\delta M_1$ with center 
$y_1(x)$ and width $\Delta_y$;
the second factor, $p(x|m_{1,2})$ occurring in eqs.
(\ref{dotm1=}) and (\ref{dotm2=}), is a Gaussian in $x$ with center 
$x_0$ and width $\Delta_x$. Defining $\mu_1= {\Ht}/{\Kt}-m_1$ with 
\BEA
\label{Ht=}
 \Ht=H+\frac{JL}{w+T/2};\qquad \Kt= K-\frac{J^2}{w+T/2},
\EEA
the parameters of these Gaussians take the form
\BEA \label{x0=}
&x_0= \Delta^2\Kt/2,\qquad \qquad&\Delta_x=\Delta^2\Kt T_e, \\
&y_1=\beta_e\mu_1\,(x_0-x), \qquad
& \Delta_y=\Delta^2\,(1-\beta_e\Kt\mu_1^2).
\EEA
Here $T_e=1/\beta_e$ is a short hand
for $T_e=\Kt(m_2-m_1^2)$; later we shall interpret
it as an effective temperature.

We are still free to choose the variance $\Delta^2$ of the 
Monte Carlo moves. To model glassy behavior we shall consider a 
dynamics that constrains the system's phase space to  
$m_2-m_1^2\ge m_0$ for some fixed  $m_0$.
This mimics a glass that has no crystalline state,
as occurs in atactic polymer melts.
For some fixed $m_0$ and $B$, we thus let $\Delta$ 
depend on the instantaneous values of $m_{1,2}(t)$ according to
\BEQ 
\label{Delta=} \Delta^2(t)=\frac{8T_e(t)}{\Kt(t)}\, \Gamma(t)
\EEQ
where
\BEQ \label{Gamma=} 
\Gamma(t)\equiv \left(\frac{B}{\mu_2(t)}\right)^{\gamma};
\quad \mu_2(t)=m_2(t)-m_1^2(t)-m_0 \EEQ
The power $\gamma$ remains a free parameter, because we will 
not make a connection to an underlying microscopy. 

The fact that $\Delta^2$ becomes large near $\mu_2=0$ implies that
there the trial moves are typically large, and thus unfavorable. 
Most of them will be rejected, 
implying, as shown below, a VFTH-type equilibrium relaxation time
\BEQ\label{taueq=}
\tau_{eq}(T)\sim \exp \Gamma _{eq}
\sim \exp\left(\frac{A(T_0)}{T-T_0}\right)^\gamma
\EEQ
while $\tau_{eq}=\infty$ for $T<T_0$. 
Though the physics of our model (\ref{H1=}) is simple, 
the choice (\ref{Delta=}), (\ref{Gamma=}) for the timescale 
allows us to derive aspects of the long time aging regime below $T_K$, 
which could well be model independent.

Let us first look at the static regime, that has $\mu_1=0$.
For large enough $T$, the variable $\mu_2$ remains positive
and  the effective temperature is trivial: $T_e^{\rm eq}=T$.
A Kauzmann transition occurs at $T_K$ set by $\mu_2=0$.
Since states with $\mu_2<0$ cannot be reached, it is evident 
that the generally expected equality $T_0=T_K$ holds. 
In statics $\mu_2$ vanishes for all $T<T_0$, implying that
$T_e^{\rm eq}$ follows from
\BEQ \frac{Km_0-T_e^{\rm eq}+T}{Km_0-T_e^{\rm eq}}
+\frac{D^2m_0}{(JT_e^{\rm eq})^2}
=\frac{J^2m_0}{(Km_0-T_e^{\rm eq})^2}\EEQ
where $D=HJ+KL$. The Kauzmann temperature is the solution of
$T_e^{\rm eq}(T_0)=T_0$. Below $T_0$, $T_e^{\rm eq}$ exceeds $T$, and
it remains finite for $T\to 0$. 
All static observables can now be expressed in $T_e^{\rm eq}$.
Since $w=[{J^2m_0+(DJ/T_e^{\rm eq})^2+T^2/4}]^{1/2}$,
we have the result for $m_1=\Ht/\Kt$ via (\ref{Ht=}), 
and using $m_2=m_1^2+m_0$,  we can find the energy from (\ref{u=}).
Since $\d T_e^{\rm eq}/\d T<1$ at $T_0^-$, the specific heat
makes an upward jump as function  of $T$; this happens also
for $-\p m_1/\p T$. The very same features occur in
the mean field glass model, the $p$-spin model,
where $T_e^{\rm eq}=x(T)/T$, with $x$ the Parisi 
breakpoint~\cite{CS}.

We shall restrict ourselves to aging experiments at  fixed $T,H$,
where the system evolves from some initial state. An example is
fast quenching from an equilibrium state 
at large temperature, for which initially $\mu_1=0$, $0<\mu_2\ll 1$. 
In the long time regime eqs. (\ref{dotm1=}), (\ref{dotm2=}) 
lead to first order differential equations for $\mu_{1,2}$ 
\BEA \label{dotmu1}
\dot \mu_1&=& 4[aJrT_e-(1+aD)(\Gamma+r)\mu_1]I \\ 
\dot\mu_2 &=& -8[\frac{rT_e}{\Kt} -(\Gamma+r)\mu_1^2]I
\label{dotmu2} \EEA
where we denote  $r=(T_e-T)/(2T_e-T)$,
$I=\int \d x\,p\,W$, and we introduce the shorthands
\BEA a=\frac{DJ^2}{\Kt^3w(w+T/2)^2},\qquad
 b=\frac{J^4T_e}{2\Kt^2w(w+T/2)^2}\EEA
$\dot\mu_2$ in eq. (\ref{dotmu2}) is dominated by the small 
factor $I\sim \exp(-\Gamma)$. 
This is the dynamic imprint of the static VFTH law (\ref{taueq=}),
which may be more general. The solution follows readily, 
since the integral of an exponential is basically the same exponential. 
It is this property that leads to the asymptotic decoupling
of decades, discussed in the introduction. 
In the aging regime  one gets $t\sim t_0\exp\Gamma$, equivalent to
\BEQ\label{mu2=}
\mu_2=\frac{B}{\left[{\ln(t/t_0)}\right]^{1/\gamma}}
\EEQ
where $t_0$ depends logarithmically on $t$. Such logarithmic
behavior is supposed to be the universal finger print for the 
glassy state; notice, however, that there can be a non-trivial
exponent. For $T>T_0$ $\mu_2$
will stick at $\mu_2^{\rm eq}\sim T-T_0$, whereas for $T<T_0$ it will 
ultimately vanish. 

To solve for $\mu_1(t)$ we
divide eq. (\ref{dotmu1}) by (\ref{dotmu2}), which 
allows to express $\mu_1$ in terms of $\mu_2$ of eq. (\ref{mu2=}),
\BEA \label{dmu1dmu2=}
\frac{\d\mu_1}{\d \mu_2}=\frac{-aJ\Kt rT_e+(1+aD)(\Gamma+r)\Kt\mu_1}
{2rT_e-2(\Gamma+r)\Kt\mu_1^2}\EEA
There exist several ranges of solutions:

$\bullet$ Regime 0:  $T >T_0$, all $\gamma$: 

Here $r\sim T_e-T\sim \mu_2$. The solution is obtained by
equating the numerator of the right hand side to zero, 
\BEQ\label{muT>T0}
 \mu_1=\frac{aJ\Kt(1+b)}{1+aD}\,\frac{\mu_2}{\Gamma}
\EEQ
Since $\mu_1\ll\mu_2$, it can be neglected to leading order.

$\bullet$ Regime 1:  $T < T_0$, $\gamma>1$: 

As in  Regime 0, but now $r$ is of order unity,
\BEQ \mu_1= \frac{aJT_e}{1+aD}\,\frac{r}{\Gamma}
\label{mu1Regime2}
\EEQ
Also in this case $\mu_1\ll\mu_2$.

$\bullet$ Regime 2: $T<T_0$, $\gamma=1$:

This is the case with a true VTFH-law.  Quite surprisingly,
eq. (\ref{dmu1dmu2=}) exhibits several subregimes with
different behavior: $\gamma=1$ is the most difficult situation,
too intricate to explain here in detail.
 
$\bullet$ Regime 3: $T < T_0$, $\gamma<1$ 

Now the denominator of (\ref{dmu1dmu2=}) becomes subleading, of order
$\Gamma\mu_2=B^\gamma \mu_2^{1-\gamma}\ll 1$, but it remains
positive, thus assuring that $\mu_2$ finally vanishes. It follows that
\BEQ \mu_1=\sqrt{\frac{rT_e}{\Kt\,\Gamma}}
(1-\frac{(1+aD)\Kt}{2\gamma rT_e}\Gamma\mu_2)\sim \mu_2^{\gamma/2}, \EEQ
so in this regime $\mu_2\ll \mu_1$ can be neglected.

The magnetization approaches equilibrium as
\BEQ
m_1(t)=\frac{\Ht^{\rm eq}}{\Kt^{\rm eq}}-\frac{1}{1+aD}\mu_1(t)
-\frac{aJ\Kt}{2(1+aD)}\mu_2(t)
\EEQ
In regimes 0 and 1 its time-dependence is dominated by 
$\mu_2 \sim (\ln t)^{-1/\gamma}$, while in regime 3 the dominant
term is $\mu_1\sim  (\ln t)^{-1/2}$.
In all cases both $m_2=m_0+m_1^2+\mu_2$ and the energy, 
see eq. (\ref{u=}), behave similar to $m_1$.

It has long been a challenge to formulate glassy behavior in 
terms of the laws of thermodynamics, because that could
point at universal behavior. Attempts to do so
remained unfruitful till recently~\cite{McKenna}\cite{Angell}.
In a recent series of papers we have shown that it can be done in
model systems where only one extra variable, the effective
temperature $T_e$, is needed to describe the physics~\cite{NEhren}
\cite{Nthermo}\cite{Nlongthermo}. The crucial step was to explain
the paradoxes concerning the Ehrenfest relations and the
Prigogine-Defay ratio~\cite{NEhren}. 
In the present model there are two independent observables, 
$m_{1,2}(t)$  or, equivalently, $m_1(t)$ and $u(t)$.
They result from the dynamics, but they can also 
be described by a thermodynamics
at an effective temperature $T_e(t)$ and an effective field $H_e(t)$. 
To do so, let us  calculate in a quasi-stationary approach 
the partition sum over macroscopically equivalent states
\BEQ
Z_e=\int Dx\,e^{-\beta_e [\H_{\rm eff}+HM_1- H_eM_1]}
\delta(M_{1,2}-Nm_{1,2}(t)) 
\EEQ 
By the extra terms
in the exponent we eliminate $H$ in favor of $H_e$.
We find for the free energy $F=-T_e\ln Z_e$ 
\BEA\label{FTTe}
F=U+(H- H_e)M_1-T\S-T_e\I 
\EEA
where $U$ and $\S$ were defined in (\ref{u=}) and (\ref{Sep=}).
The new term  $\I=N[1+\ln(m_2-m_1^2)]/2$ is the 
entropy of the configurations with given $m_{1,2}$. Though our model is 
classical, we can impose $\I=0$ at $T=0$ by adding a constant,
yielding  
$ \I=(N/2)\ln(1+\mu_2/m_0)$.
In his seminal paper Kauzmann pointed out that  
the equilibrium configurational entropy vanishes already at $T_K$.
We see that this is indeed the case here.
We can also check another aspect of the standard folklore.
Comparison with eq. (\ref{taueq=}) yields
$\ln \tau_{eq}$$\sim$ $(N/\I)^\gamma$. This is the generalization
of the Adam-Gibbs relation~\cite{AdamGibbs} to cases with $\gamma\neq1$.

$T_e$ and $H_e$ are required to satisfy the saddle point equations that
follow from varying $m_{1,2}$ in eq. (\ref{FTTe}), implying
\BEA\label{Tedyn=} 
T_e&=&\Kt(m_2-m_1^2)=\Kt(m_0+\mu_2)\\  
\label{Hedyn=}
H_e&=&K m_1-\frac{J(Jm_1+L)}{w+\hT} =H-\Kt\mu_1
\EEA
They only depend on $H$ through $m_{1,2}=m_{1,2}(t;T,H)$.

In eqs. (\ref{x0=})-(\ref{Delta=}) it was already seen 
that the expression (\ref{Tedyn=}) for $T_e$ was useful 
to shorten the dynamical equations.
 It can now be checked that $T_e-T_e^{\rm eq}
\sim +\mu_2-\mu_1$, so it has no definite sign, whereas 
$H_e-H\sim-\mu_1$ is negative. 
It is now seen that the ``simple'' Regimes 0 and 1 have a natural
interpretation: one then has $\mu_1\ll \mu_2$, 
so the effective field $H_e\approx H$  is not needed:
the glassy state is coded by the effective temperature alone.
Such a situation was encountered before in model glasses with $T_0=0$
~\cite{Nhammer}\cite{Nlongthermo}, and in the $p$-spin model
~\cite{NEhren}\cite{HSN}.
In the ``complicated'' Regimes 2 and 3 
one ultimately has $T_e-T_e^{\rm eq}\sim H_e-H$. Both parameters 
are thus needed to describe the evolution, a result
that should be valid beyond our model.
Systems with a true VFTH-law ($\gamma=1$) are at the borderline,
and thus more subtle.

We now consider which processes are involved at large time $t$.
The change of energy may be cast in the form
\BEQ \dot u(t)=-\frac{u(t)}{\tau(T,T_e(t))}\EEQ
For $T>T_0$ the behavior $I \sim \exp(-\Gamma)$ 
reproduces at equilibrium eq. (\ref{taueq=}); 
Outside equilibrium it follows that $\tau(T,T_e)\approx \tau_{eq}(T_e)$, 
as was observed in systems with an Arrhenius law~\cite{Nhammer}.
For $T<T_0$, $\gamma>1$ (Regime 1) 
the timescale of a system characterized by $T$ and $T_e$ 
 obeys a generalization of the VFTH-law
\BEQ \tau(T,T_e)\sim \exp\Gamma=
\exp\left(\frac{A(T)}{T_e-T_e^{\rm eq}(T)}\right)^\gamma.
\EEQ
saying that the equilibrium VFTH law has a definite fingerprint on the
aging dynamics below $T_0$.
Here, as well as in eq. (\ref{taueq=}), $A(T)=B\Kt (1+b/(1+aD))$.
In the cases $\gamma\le 1$ there are various behaviors, 
that may not be universal.
In all these cases it holds that for $t\gg t_0$ 
the active  processes are indeed those with timescale 
$\tau(T,T_e(t))\sim t$, as is consistent with the logarithmic
time dependence: $u(t)=\tilde u(\ln t)$ implies 
$\dot u=-u/\tau$ $\to$ $\tau=(-u/\tilde u')\,t$.

Let us now consider the results from a more general thermodynamic 
point of view. The first law says that the change in energy 
 is equal to the heat added to the system and the work done on the
system, $\d U=\dbarrm Q-M_1\d H$. It is satisfied with 
\BEA \label{dQ=}\dbarrm Q&=&T\d \S\,+\,T_e\d\I+(H_e-H)\d M_1 
\EEA
This holds because of the presence  of a free energy, 
eq. (\ref{FTTe}), satisfying $\d F=-\S\d T-\I\d T_e-M_1\d H_e$.
It does not invoke the relation between $m_1(t)$ and $m_2(t)$, 
which differs in the various regimes. 
The two entropic terms in (\ref{dQ=}) are related to the 
fast and slow processes, respectively~\cite{Nthermo}\cite{NEhren}. 
The last term is new; it results from irreversible internal work.

In the aging process
the system will go to lower energy by producing heat at
a rate $|\dbarrm Q/\d t|$, that follows from (\ref{dQ=}).
It holds that $ |\dbarrm Q/\d t|\sim |\dot \mu_1|+|\dot \mu_2|$ .
The dominant term is $\dot \S+\dot \I\sim \dot \mu_2$ in regimes 0 and
1, while it is solely due to $\dot \S\sim \dot \mu_1$ in regime 3. 

In thermodynamics close to equilibrium a central role is played by 
the irreversible entropy production. For systems far from
equilibrium, this is also the case.
We now show that it can be derived explicitly for systems
such as the present one. We write the change in total entropy $S=\S+\I$
as $\d S=\d_e S+\d_i S$,
where $\d_e S =\dbarrm Q/T $ is the externally supplied entropy,
and $\d_i S$ the internally produced entropy.
This gives the general result
\BEQ \label{Sprod}
\frac{\d S_i}{\d t}=\frac{T-T_e}{T}\,
\frac{\d \I}{\d t} +\frac{H-H_e}{T}\,\frac{\d M_1}{\d t}
\EEQ
It can be verified explicitly in our model that both contributions
are non-negative, in accordance with the second law.
The first term of (\ref{Sprod}) is of order $r\dot \mu_2$, and is
dominant in regimes 0 and 1; The last term is
of order $\mu_1|\dot\mu_1|$, and it dominates in regime 3.

In conclusion, we introduce an exactly solvable model glass
with a set of fast variables (spherical spins) and a set of slow
variables (harmonic oscillators). The fast modes are summed out.
For the slow modes we assume a dynamics that obeys a 
generalized VFTH law, $\tau_{eq}\sim \exp\,A^\gamma(T-T_0)^{-\gamma}$.
In statics there occurs a Kauzmann transition.
We  solve the aging dynamics, and 
verify a generalized Adam-Gibbs relation between the
relaxation time and the configurational entropy.
Below $T_0$ the relaxation time for a system described by its effective
temperature may have a VFTH form, or a different one.
The dynamically active processes have timescale $\sim t$.

The aging solution can be described by a quasi-static 
thermodynamic approach.
Above the Kauzmann temperature $T_0$
one extra variable is needed, the effective temperature, that depends
logarithmically on time. Below $T_0$ this persists in some cases,
while in other cases also an effective field is needed.
Systems with the original VFTH-law are at the borderline.

We finally derive the production rates of entropy and heat for systems
described by an effective temperature and field.
They are non-vanishing whenever the effective 
temperature deviates from the bath temperature, and/or
when the effective field deviates from the external field. 

%\acknowledgments
We thank  A. Allahverdyan, A. Crisanti, L. Leuzzi, L.S. Suttorp
and G.H. Wegdam for discussion.

\end{multicols}
\end{document}